\documentclass[vecphys]{svmult}

\usepackage{makeidx}         % allows index generation
\usepackage{graphicx}        % standard LaTeX graphics tool
\usepackage{multicol}        % used for the two-column index
\usepackage[bottom]{footmisc}% places footnotes at page bottom
\usepackage{rotating}
\makeindex             % used for the subject index
\newcommand{\snia}{SN\,Ia }
\newcommand{\snii}{SN\,II }
\newcommand{\aca}{AcA}
\newcommand{\apj}{ApJ}
\newcommand{\apjs}{ApJS}
\newcommand{\apjl}{ApJ}
\newcommand{\aj}{AJ}
\newcommand{\mnras}{MNRAS}
\newcommand{\pasp}{PASP}
\newcommand{\aap}{A\&A}
\newcommand{\aaps}{A\&AS}
\newcommand{\apss}{Ap\&SS}
\newcommand{\araa}{ARA\&A}

\begin{document}

\title*{Horizontal Branch Stars\\and the Ultraviolet Universe}
\titlerunning{HB Stars and the UV Universe}
\author{M. Catelan}
\institute{Pontificia Universidad Cat\'olica de Chile, Departamento de Astronom\'ia y 
  Astrof\'isica, Av. Vicu\~na Mackenna 4860, 782-0436 Macul, Santiago, Chile  
\texttt{mcatelan@astro.puc.cl}}
\maketitle

\begin{abstract}
Extremely hot horizontal branch (HB) stars and their progeny
are widely considered to be responsible
for the ``ultraviolet upturn'' (or UVX) phenomenon observed in elliptical galaxies
and the bulges of spirals. Yet, the precise evolutionary channels that lead 
to the production of these stars remain the source of much
debate. In this review, we discuss two key physical ingredients that are 
required in order for reliable quantitative models of the UV output of stellar 
populations to be computed, namely, the mass loss rates of red giant branch 
stars and the helium enrichment ``law'' at high metallicities. 
In particular, the recent evidence pointing towards a strong enhancement 
in the abundances of the $\alpha$-elements in the Galactic bulge (compared 
to the disk), and also the available indications of a similar overabundance 
in (massive) elliptical galaxies, strongly suggest that the helium abundance 
$Y$ may be higher in ellipticals and bulges than it is in spiral disks by an 
amount that may reach up to 0.15 at ${\rm [Fe/H]} \sim +0.5$. If so, this would 
strongly favor the production of hot HB stars at high metallicity in galactic 
spheroids. We also discuss the existence of mass loss recipes beyond the commonly 
adopted Reimers ``law'' that are not only more consistent with the available 
empirical data, but also much more favorable to the production of extended HB 
stars at high metallicity. Finally, we discuss new empirical evidence that 
suggests that different evolutionary channels may be responsible for the 
production of EHB stars in the field and in clusters. 
\end{abstract}

\section{Introduction}
\label{sec:1}
Horizontal branch (HB) stars are the immediate progeny of low-mass red 
giant branch (RGB) stars. They appear to have been discovered by \cite{tb27} 
in an analysis of data collected by \cite{hs15} for the globular cluster M3 
(NGC~5272), and were first correctly identified as stars that burn helium 
in their cores and hydrogen in a shell by \cite{hs55}. Recent reviews 
dealing with the general properties and astrophysical importance of these 
stars have been provided in \cite{sm04,mc07b}, whereas a recent overview of 
low-mass stellar evolution can be found in \cite{mc07a}. 

For a given chemical composition, and after losing mass on the RGB, such 
stars end up at different positions along the zero-age HB (ZAHB), depending 
on the total amount of mass lost during their RGB ascent (e.g., 
\cite{vcea69,ir70,jf72,br73}). Thus, depending on the total mass loss, RGB 
stars of a given chemical composition can become cool red HB stars, very hot 
(and thus UV-bright) extended (or extreme) HB (EHB) stars (\cite{gs74})~-- 
or a whole ``rainbow'' of intermediate options, including the RR Lyrae 
(pulsating) stars and the regular (A- and B-type) blue HB stars. Blue 
subdwarf (sdB) stars have long been identified as the field counterparts 
of the EHB stars that are found in globular clusters (\cite{gs74}). EHB status 
may also characterize at least some among the cooler O-type subdwarf (sdO) 
stars, whereas other sdO's likely represent the progeny of EHB stars 
(\cite{uhea06}). Among the UV-bright progeny of hot HB stars one also 
finds the so-called ``asymptotic giant branch (AGB) manqu\`e''  stars 
(\cite{gr90}) and the ``post-early AGB'' stars (\cite{ebea90}), whose 
evolutionary properties are further described in \cite{bdea93,mc07a}.

In general, for a given chemical composition, larger 
amounts of mass loss lead to bluer positions on the ZAHB, and thus to  
more efficient far-UV emitters. In fact, even beyond the EHB proper 
there may still be ``life'' on the HB phase: 
as recently discussed by several authors, the RGB
progenitors of HB stars may (somehow) lose so much mass prior to arriving 
on the ZAHB that they may miss the helium ``flash'' at the RGB tip 
altogether, but still ignite helium during the white dwarf cooling 
curve (e.g., \cite{tbea01,scea03}). Such ``late flashers'' are predicted 
to be even hotter than EHB stars, and have quite anomalous surface 
abundances compared to EHB stars (see, e.g., \cite{smea02,smea04,tlea04}). 
In the observed color-magnitude diagrams (CMD's) of Galactic globular 
clusters, such stars have been identified as the ``blue hook'' feature 
that is seen towards the very hot end of blue HB ``tails'' (e.g., 
\cite{dcea96,dcea00,jwea98,area04,vrea07}). 

For a given chemical composition, 
bolometric luminosity is roughly constant along the ZAHB. Therefore, the 
hotter a star becomes when it reaches the ZAHB, the higher its potential 
contribution to the population's total UV output. As a consequence, in 
order to be able to reliably predict the UV light emanating from a  
given stellar population that may contain HB stars (i.e., in which a 
sufficiently old component is present), one must be able to reliably 
predict the distribution of temperatures along the HB. 

This is far from being a trivial task. Such a temperature distribution, 
for a given chemical composition, is mainly determined by mass loss on the 
RGB (in addition to age). Therefore, one must accordingly be able to reliably 
compute mass loss rates (and how they vary with time) 
for RGB stars in order to be able to predict the 
UV colors of such a stellar population. 

In addition to mass loss and age effects, both theoretical and empirical 
evidence reveals that the temperature of a star along the HB depends on 
its chemical composition, as indicated by its metallicity $Z$ 
and helium abundance $Y$. The former plays the well-known role of ``first 
parameter'' in the determination of HB morphology, whereas the latter, 
which was the first suggested ``second parameter,'' has recently regained  
popularity as a second-parameter candidate (see \cite{mc07b} for a 
recent review). In short, HB type tends to become redder with increasing $Z$, 
whereas it tends to become bluer with increasing $Y$. Therefore, in 
order to be able to reliably predict the UV output from an old stellar 
population, one needs not only accurate mass loss rates, but also to 
know the variation in the helium abundance with metallicity. 

From 
these basic conclusions to the scenarii in which very old, metal-poor stellar 
populations (e.g., \cite{ywl94,pl97}), or old, metal-rich, but very helium-enriched 
populations (e.g., \cite{bdea95}), are responsible for the UV upturn in 
the observed spectra of elliptical galaxies and bulges of spirals (see the 
several excellent reviews in this volume) is just a fairly immediate logical 
step. Evidence in favor of the high-metallicity scenario is provided by the 
presence of EHB stars in high-$Z$  
{\em open} clusters, including the $\sim 7$~Gyr-old (\cite{asea99})
NGC~188 (\cite{wlea98}) and the $\sim 8$~Gyr-old (\cite{gcea06}) NGC~6791 
(\cite{ku92,kr95,wlea98,lbea06,gcea06,egea06}). Both of these clusters are 
well known for their supersolar metallicities (e.g., \cite{wj03}).

It is at present quite well established that the UV upturn is indeed 
due to hot HB stars and their progeny (e.g., \cite{gr90,roc99,tb04,crea07}). 
Direct confirmation 
that such stars are present in the cores of elliptical galaxies, and 
thus are most likely responsible for the UV upturn phenomenon, has been 
provided by the deep {\em Hubble Space Telescope} near-UV STIS images of 
the nearby elliptical galaxy M32 (see \cite{tbea00}). In addition, such stars 
have also been found in our own Galactic bulge (\cite{gbea96,gbea05}). 
Therefore, if viewed from an outside galaxy, our own Milky Way's bulge 
would also present a UV upturn~-- and it would be caused by the presence 
of EHB stars (and their progeny).  

The main question that remains unanswered, therefore, is: {\em what is the 
physical origin of these hot HB stars?} In other words, is the UV upturn 
phenomenon due to very old, metal-poor stellar populations, or is it due 
instead to highly helium-enriched metal-rich populations? A third possibility 
is that binary star evolution is the actual culprit, as recently discussed 
by \cite{zhea07} (see also Sect.~\ref{sec:channels} for further discussion).  
In this review, our main goal is not to provide a solution to this 
long-standing problem. Instead, we shall focus on some key ingredients 
that, in our view, must 
be better treated in the theoretical models if we ever wish to be able to 
obtain convincing predictions of the UV output of a given (old) population. 
While our main focus is on single-star evolution, the caveats we 
raise should also be of relevance in the case of binaries.

\begin{sidewaysfigure}[!tp]
  \includegraphics[scale=0.6]{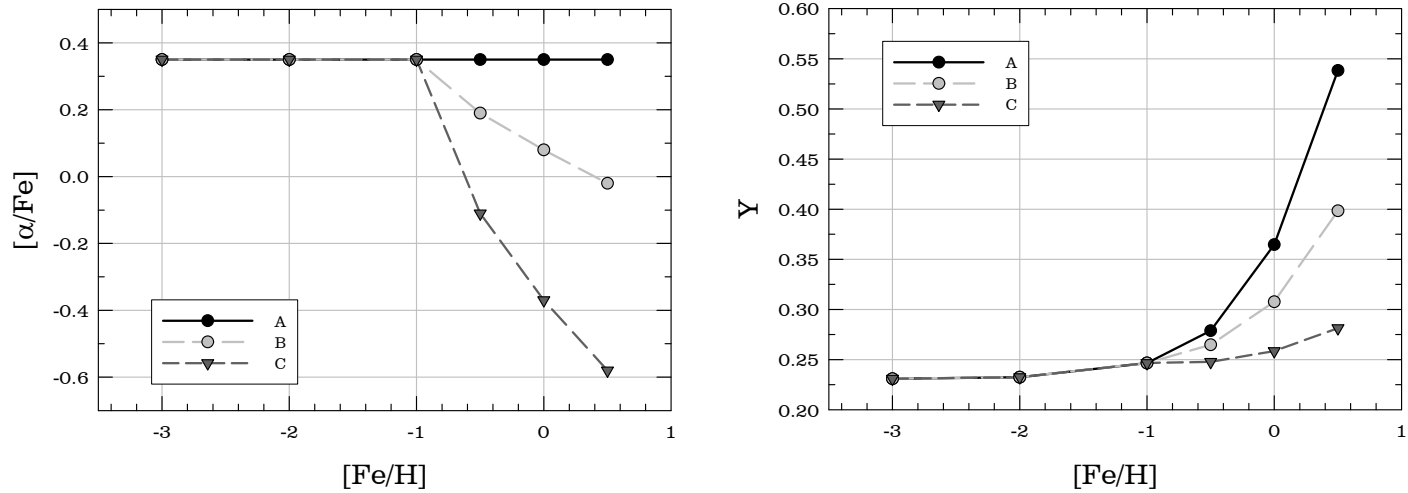} 
  \caption{The results of simple chemical evolution models (\cite{cfp96}) 
    are shown in the $[\alpha/{\rm Fe}] - [{\rm Fe/H}]$ plane ({\em left})   
    and in the corresponding helium abundance $Y - [{\rm Fe/H}]$ plane 
    ({\em right}) for three different scenarios regarding the maximum relative 
    fraction of \snia events: 0\% (Case~A), 10\% (Case~B), 40\% (Case~C). 
    As can clearly be seen, high-metallicity stellar populations with
    a strong level of $\alpha$-element enhancement are expected to be 
    strongly enriched in helium as well.   
  \label{fig:chemevol}
}
\end{sidewaysfigure}

\begin{figure}[!tp]
  \includegraphics[angle=0,scale=0.76]{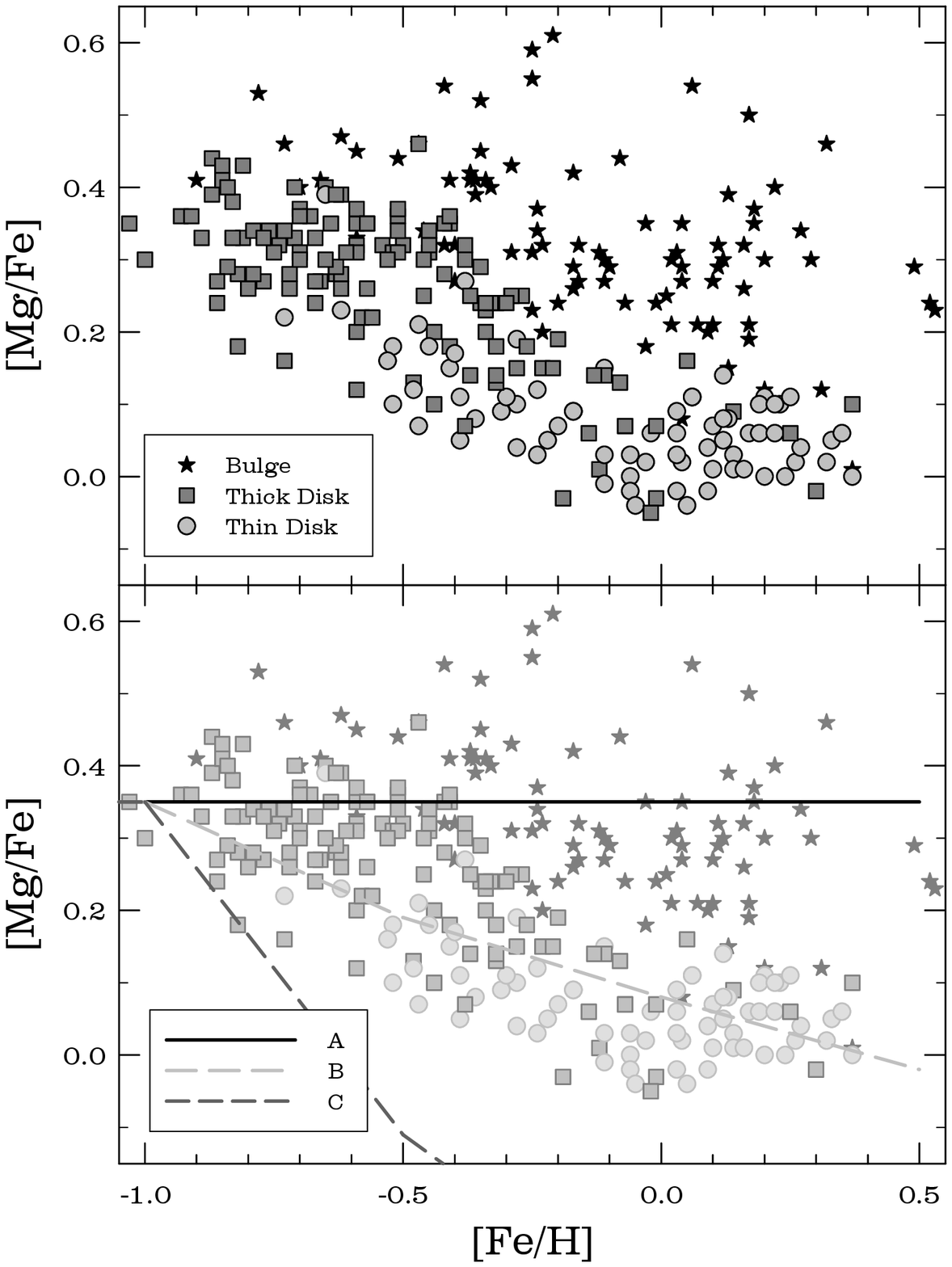} 
  \caption{{\em Upper panel:} 
    Observed Mg abundances (taken as representative of the 
    abundances of the $\alpha$-capture elements) as a function of 
    [Fe/H] for Galactic bulge ({\em star symbols}), thick disk
    ({\em squares}), and thin disk ({\em circles}) stars. As can 
    clearly be seen, bulge stars are enhanced in Mg relative to 
    disk stars. {\em Lower panel:} same as in the upper panel, 
    but with the chemical evolution models shown in Figure~\ref{fig:chemevol}
    overplotted. ``Case~A'' models, in which \snii play the 
    most important role, provide a better description of the bulge 
    stars than do the other chemical enrichment scenarios, thus 
    implying a large helium enhancement in high-metallicity bulge 
    stars (see the right panel in Figure~\ref{fig:chemevol}) as
    compared to disk stars with similar [Fe/H].  
  \label{fig:bulgedisk}
}
\end{figure}

\section{The Helium Enrichment ``Law'' in Different Stellar Populations}
\label{sec:helium}

It is now very well established that, even within one individual galaxy,  
different stellar populations follow different chemical enrichment ``laws,'' 
as indicated by the measured abundances of the so-called $\alpha$-capture 
(including Ne, Mg, Si, S, Ca, Ti), $r-$process (e.g., Rh, Ag, Eu, Pt), and 
$s-$process (e.g., Sr, Zr, Ba, La) elements. This becomes particularly 
evident when analyzing their respective trends with [Fe/H] 
(e.g., \cite{cwea89,bpea05}). 

Recently, it has become particularly clear 
that the Galactic bulge follows a different enrichment law than does the 
Galactic thick disk, which in turn follows a different law from the 
thin disk (e.g., \cite{mzea06,cs06,sbea07,alea07,jfea07,rmrea07}). As pointed 
out by several of the quoted authors, the overabundance in at least some of 
the $\alpha$-capture elements up to high metallicities for the bulge 
stars likely indicates a very short formation timescale for the Galactic 
bulge, whose chemical enrichment seems accordingly to have been dominated 
by ejecta from type II supernova (SN\,II) explosions. In like vein, (massive) 
elliptical galaxies also appear to be characterized by high metallicities 
and supersolar $\alpha$-to-iron ratios (e.g., 
\cite{gwea92,rdea93,sb95,lg97,dmea03,amea07}). 

What is the relation between these well-known trends and the production
of hot HB stars in galaxies? Here we would like to emphasize the often 
overlooked fact that different $\alpha$-enrichment trends with [Fe/H] 
are also expected to be accompained by different trends in the helium 
enrichment ``law'' with [Fe/H]. This is clearly shown in 
Figure~\ref{fig:chemevol}, which is based on the simple chemical evolution 
calculations carried out by \cite{cfp96}, in which the contribution of 
type Ia supernovae (SN\,Ia) to the chemical enrichment is assumed to increase 
with [Fe/H] above ${\rm [Fe/H]} = -1.0$, and to be negligible for lower
metallicities. In Case~B, the relative number of \snia events reaches 
a maximum of 10\%, whereas in Case~C a maximum of 40\% of \snia events 
is assumed. Case~A assumes the contribution of \snia to be negligible 
even at high metallicities. As can clearly be seen, the variation in 
the helium abundance with [Fe/H] depends very strongly on the relative 
contributions of different types of supernovae to the chemical enrichment
of a stellar population. In other words, different 
$[\alpha/{\rm Fe}] - {\rm [Fe/H]}$ curves imply different 
$Y - {\rm [Fe/H]}$ curves as well~-- and this {\em must} be accounted 
for in the detailed modelling of $\alpha$-enhanced, high-metallicity 
stellar populations, such as appears to be the case in elliptical 
galaxies and the bulges of spirals.  

Figure~\ref{fig:bulgedisk} ({\em upper panel}) shows the available Mg-to-Fe 
abundance ratios for Galactic bulge stars ({\em star symbols}), compared 
to similar data for Galactic thick ({\em squares}) and thin 
({\em circles}) disk stars. Data for bulge stars were taken from the 
studies by \cite{sbea07,alea07,rmrea07}, which were complemented with 
data for disk stars from \cite{tbea05,brea06}. This plot clearly 
confirms the already well-known differences in $\alpha$-enhancement 
behavior for the different metal-rich Galactic components. The 
{\em lower panel} in Figure~\ref{fig:bulgedisk} overplots the models by 
\cite{cfp96} previously shown in Figure~\ref{fig:chemevol} on these data. 
This plot clearly indicates that a high contribution of \snii is needed 
to explain the high $\alpha$-enhancement levels observed among bulge 
stars. Figure~\ref{fig:chemevol}, in turn, reveals that such a high \snii 
contribution also implies that high-metallicity stars in the Galactic bulge 
are likely to be He-enhanced compared to disk stars with similar [Fe/H], 
by an amount which can reach almost 
$\Delta Y_{{\rm bulge}-{\rm disk}} \approx 0.15$ at 
${\rm [Fe/H]} = +0.5$. 

As a consequence, 
such a high level of $\alpha$-enrichment can have very important 
consequences for the production of extremely hot HB stars in elliptical 
galaxies and the bulges of spirals. As shown by several different authors, 
including \cite{ehea92,abea94,bdea95,syea97}, for the observed UV upturn 
in the observed spectra of these galaxies to be successfully accounted 
for in terms of single-star, high-metallicity evolutionary models, one 
needs a very high level of helium enrichment. The available spectroscopic 
evidence, by pointing towards high $\alpha$-enrichment levels being present 
in these galaxies, may provide a natural explanation for the required 
(high) helium abundances. 
In any case, we caution that a reliable, quantitative assessment of the 
level of helium enrichment to be expected in different populations 
associated with galaxies of different Hubble types will require 
considerable progress to be achieved in the modelling of the yields of 
high-mass stars, the shape of the initial mass function (especially at
its high-mass end), and in our understanding of the very origin and 
formation history of these galaxies (e.g., \cite{sbea07,jfea07,fm07}).

\section{The Mass Loss ``Law'' for RGB Stars}
\label{sec:mloss}

The amount of 
mass lost by an RGB star is crucial in determining the temperature 
such a star will end up with when it reaches the ZAHB 
(e.g., \cite{br73,bdea93}). In particular, in order 
to become EHB stars, the progenitors of HB stars must lose 
large amounts of mass during the RGB phase. Therefore, in order 
to reliably predict the UV output from a stellar population, 
reliable mass loss rates for low-mass RGB stars are required. 

Unfortunately, we remain unable at present to compute reliable mass 
loss rates for red giants, either from detailed theoretical models 
based on first physical principles~-- which are still lacking in the 
literature~-- or from semi-empirical recipes based on observational 
material. 

This is not to mean that such semi-empirical recipes are also lacking; 
quite the contrary, in fact: as discussed in \cite{mc00,mc07b}, there 
exist at present several different such recipes which are equally 
satisfactory at describing the available mass loss rates for red 
giant stars. Some such recipes are given in Figure~\ref{fig:equations}
(see the Appendix in \cite{mc00} for more details). 
All these formulae are based on 
original expressions suggested by the indicated authors; thus 
``modified Reimers'' is reminiscent of the original Reimers mass 
loss formula (\cite{dr75a,dr75b}), but allowing the exponent on the 
right-hand side of the equation to be determined by the data, as 
opposed to being imposed a priori. The other expressions also have 
exponents determined by least-squares fits to the data~-- and these 
exponents generally differ somewhat from the originally proposed 
ones (for the Mullan expression, see \cite{dm78}; Goldberg's recipe 
appears in \cite{lg79}; Judge \& Stencel's is given in \cite{js91}; 
finally, the VandenBerg expression appears first in \cite{mc00}). 

What is especially important to note here is that these expressions, 
though all 
roughly equivalent in their capacity to describe the available mass 
loss data (\cite{mc00}), 
imply integrated mass loss values that differ from one 
case to the next, thus also implying different metallicity and 
age dependencies (see also \cite{mc00,mc07b}). This is more clearly 
shown in Figure~\ref{fig:DMFeHAge}: in the {\em upper panel}, one 
sees the different metallicity dependencies for a fixed age (9~Gyr), 
with the absolute integrated mass loss values having been normalized 
to $0.10\,M_{\odot}$ at ${\rm [Fe/H]} = -1.5$; likewise, the {\em lower
panel} shows the different age dependencies at a fixed metallicity 
(${\rm [Fe/H]} = -0.71$), where the absolute integrated mass loss 
values have been normalized 
to $0.20\,M_{\odot}$ for an age of 12~Gyr. These plots show very 
clearly that most of the available recipes predict both a metallicity 
and an age dependence that are {\em stronger} than in the case of the 
commonly adopted Reimers ``law''~-- the only exception being, in fact,
the milder age dependence predicted by the Goldberg expression.

\begin{figure}[!tp]
  \includegraphics[angle=0,scale=0.57]{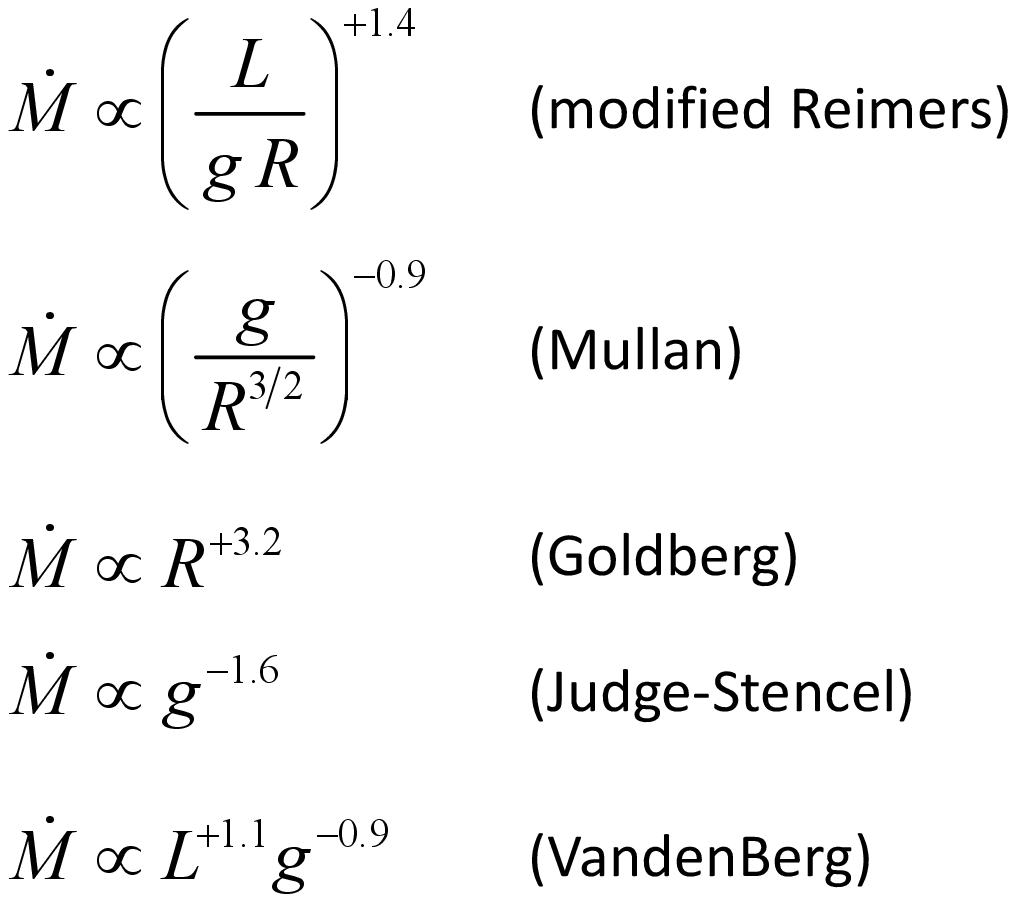} 
  \caption{Mass loss ``laws'' for red giant stars. As shown in 
   \cite{mc00}, all these expressions are equally successful in 
   accounting for the available (measured) mass loss rates for
   red giants. By contrast, the original Reimers formula 
   (\cite{dr75a,dr75b}) is unable to describe the available 
   empirical data. 
  \label{fig:equations}
}
\end{figure}

\begin{figure}[!tp]
 \includegraphics[angle=0,scale=0.65]{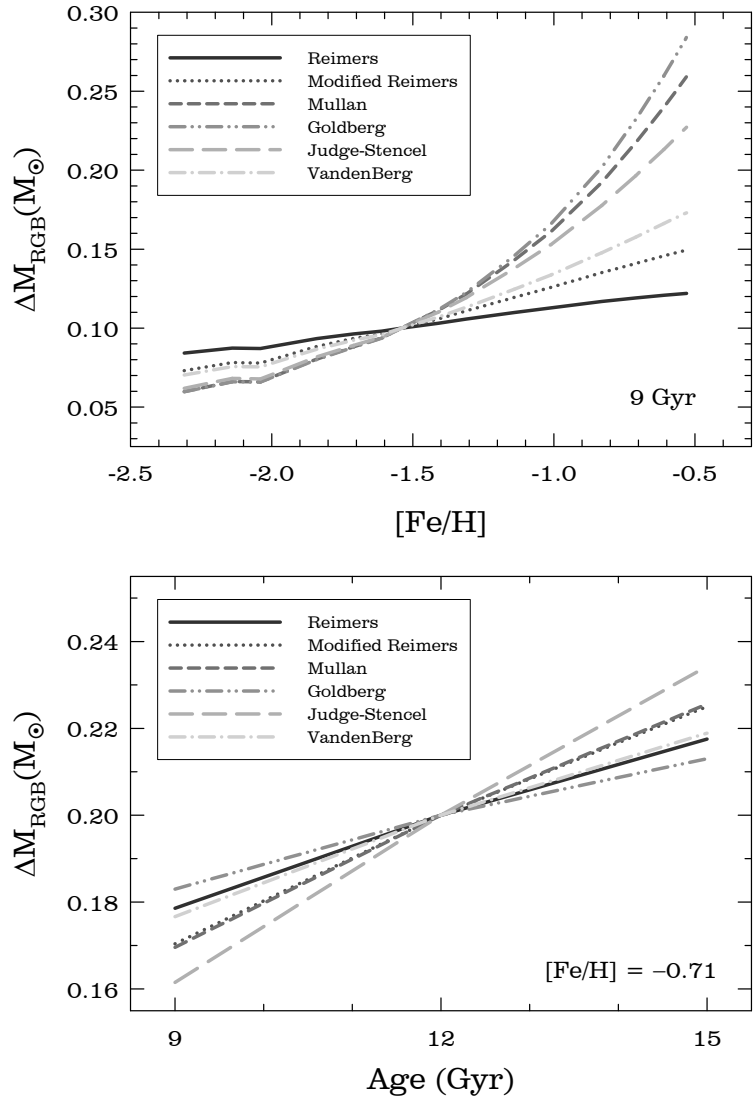} 
  \caption{Integrated RGB mass loss variation with [Fe/H] at an age 
    of 9~Gyr ({\em upper panel}) and with age at a fixed 
    ${\rm [Fe/H]} = -0.71$ ({\em lower panel}). In the upper panel, 
    mass loss has been normalized to a value of $0.10\,M_{\odot}$ 
    at ${\rm [Fe/H]} = -1.5$, whereas in the lower panel the normalization
    condition specifies an integrated mass loss of $0.20\,M_{\odot}$ 
    at 12~Gyr. The different lines in these plots 
    indicate the predicted (integrated) RGB mass loss for the several 
    different mass loss recipes provided in Figure~\ref{fig:equations}, 
    and also for the original Reimers (\cite{dr75a,dr75b}) expression. 
    Note that the mass loss variation with both metallicity and age 
    depends on the adopted recipe for $\dot{M}$.  
  \label{fig:DMFeHAge}
}
\end{figure}

Unfortunately, 
until we are in a position to reliably decide which of these several
expressions (if any; see \cite{mc00,mc07b} for several lingering
caveats) provides a better description of the available 
data, we will remain fundamentally unable to reliably predict the 
UV output from a given stellar population. A safer approach at 
present would likely be to analyze independently
the impact of each of these 
recipes upon the model predictions~-- which should give us a better 
handle of the systematic errors that are brought about by our lack 
of a unique, reliable description of mass loss rates in red giant stars. 

The (future) availability of such a description notwithstanding, the 
empirical evidence strongly suggests that mass loss would 
{\em still} retain a stochastic component, as indicated by  
the spread in colors that is {\em always} present in the 
observed CMD's of globular clusters, and which reveals the presence 
of a spread in mass loss rates at the (typical) level of 
$\sigma_M \simeq 0.02-0.03\,M_{\odot}$ (e.g., 
\cite{br73,ywl90,vdea96,vc07}). The existence of such a spread in mass 
loss is obviously very important 
in order for reliable predictions of the UV output from 
a given stellar population to be obtained, 
since it is precisely those stars at the high-$\Delta M$ 
tail of the mass loss distribution that will end up at hotter positions 
on the ZAHB, and thus produce the higher amount of UV light.
In fact, the presence of spectroscopically confirmed (\cite{smea00}) 
hot blue HB stars in such famous {\em red} HB clusters as 47~Tucanae 
(NGC~104) and NGC~362 is proof of the importance of quantitatively 
establishing the spread in mass loss rates in order to determine 
the number of UV contributors in a given stellar population. 

In this sense, while it has been suggested that such a dispersion 
in mass loss may be related to the central density of stellar systems, 
and hence most likely to 
dynamical effects (\cite{mcea01}), a systematic study is still 
lacking, and is further complicated by the discovery of multiple 
stellar components with different levels of helium enrichment in 
several among the most massive Galactic globulars (e.g., 
\cite{cd07,gpea07} and references therein). 

One way or another, it remains unclear at present the level to which 
stochastic spreads in the integrated mass loss should be adopted in 
models of elliptical galaxies and the bulges of spirals. In fact, 
the suggested dependence on central density for Galactic globular 
clusters (\cite{mcea01}) may point towards a smaller spread in mass 
loss in the field populations 
of galaxies than is seen in high-density globulars.

\section{Different Channels for EHB Stars in the Field vs. Clusters?}
\label{sec:channels}

It has recently been shown (\cite{mbea06,mbea07,mbea07b}) that (close) binary 
systems are not present in significant numbers among the hottest HB stars 
in Galactic globular clusters. In this sense, \cite{mbea07b} estimate a 
most likely (close) binary fraction $f = 4\%$ among EHB stars in NGC~6752, 
to be compared with an estimated $f = 42-69\%$ among field sdB stars 
(\cite{pmea01,rnea04,lmrea06}; see also \cite{zhea03,zhea07} for 
additional references to field star work). 

This may point to different evolutionary channels for the production 
of EHB stars in globular clusters and the field. If so, the usual 
approach of using observations of resolved globular clusters as a 
guide to the physical origin of the UV upturn phenomenon affecting 
elliptical galaxies and spiral bulges may in fact be quite inadequate. 
Note that \cite{mbea07b} suggest that the different $f$ fractions between 
field and globular cluster stars may be due to an $f-$age relation, with 
the field systems being on average significantly younger than those 
found in globulars. 

In fact, it is not unlikely that the primordial binary fraction in 
globular clusters is quite low. The CMD study by \cite{cwea08} shows 
that the main-sequence binary fraction is small away from the center 
in NGC~6752, whereas {\em Hubble Space Telescope} data suggest that it 
is in the $15-38\%$ range closer to the core (\cite{rb97}). According 
to the realistic $N$-body simulations by \cite{jhea07}, one expects 
the primordial binary frequency of a cluster to be well preserved 
outside the cluster's half-mass radius, thus supporting a small 
primordial binary fraction in NGC~6752, in line with the EHB results
by \cite{mbea06,mbea07,mbea07b}. A similar result was obtained 
very recently in the case of NGC~6397 (H. Richer, {\em priv. comm.}). 
Intriguingly, \cite{cl06} has very recently argued against a 
high primordial binary fraction among {\em field} stars as well, 
pointing out that $\sim 2/3$ of all main-sequence stellar systems 
in the Galactic disk appear to be single, and arguing that the most 
common outcome of the star formation process is a single rather than 
a multiple star. One way or another, the seemingly high primordial 
binary fraction for field sdB stars does give some support to the 
scenario in which binaries play a relevant role in explaining the 
UV upturn phenomenon (\cite{zhea07}). 

\section{Conclusions}
\label{sec:concl}

The UV upturn (or UVX) phenomenon affecting elliptical galaxies and the bulges
of spirals remains at present one of the most intriguing problems encountered 
at the interface between stellar and extragalactic astronomy. While it appears 
clear that the stars responsible for the observed far-UV light are the so-called 
EHB stars (and their progeny), it is not clear in detail how these stars are 
produced. This has given rise to several different proposed explanations
for the detected upturn in the observed far-UV spectra, 
including the ``low metallicity, very old ages'' scenario
(e.g., \cite{ywl94,pl97}), the ``high metallicity, high helium abundances'' 
scenario (e.g., \cite{ehea92,abea94,syea97}), and the binary stars scenario 
(\cite{zhea07}). Whatever one's favorite theoretical framework, 
reliable predictions cannot be obtained without adequate knowledge 
of the underlying variation in the helium abundance with metallicity 
and in the integrated RGB mass loss with both metallicity and age. 

Concerning the helium enrichment ``law,'' we have argued that the available 
spectroscopic data for Galactic field stars, which indicate different 
$\alpha$-capture enrichment laws for different stellar populations (bulge, 
thick disk, thin disk), also imply different {\em helium abundance} trends 
with [Fe/H], the resulting differences in $Y$ (at fixed [Fe/H]) potentially 
reaching very significant levels 
(i.e., $\Delta Y_{{\rm bulge}-{\rm disk}} > 0.1$) at supersolar [Fe/H]. 
Therefore, in the high metallicity scenario, it should be easier to produce 
EHB stars for a bulge chemical enrichment law than if one assumes a 
``universal,'' disk-like ``law'' to apply for different stellar populations 
in galaxies of different Hubble types. 

Hot HB stars would not exist if their immediate progenitors, namely low-mass 
RGB stars, did not lose substantial amounts of mass prior to arriving on the 
ZAHB. Unfortunately, we remain fundamentally unable to predict the 
amount of mass that a given RGB star will lose as it climbs up the RGB, 
due both to the lack of suitable theoretical models and to insufficient 
observational data constraining the phenomenon. While the original 
semi-empirical mass loss formula by Reimers (\cite{dr75a,dr75b}) has been 
shown to be inconsistent with more recently derived mass loss rates, several 
alternative formulations have been proposed which can all account for the 
more modern
observational data equally satisfactorily. This limits the extent to which 
we can build predictive models of the UV upturn phenomenon, since 
each of the available mass loss recipes predicts a different 
mass loss dependence on metallicity and age. An important 
breakthrough in our understanding of mass loss rates in red giants 
will be needed 
before we are in a position to conclusively corroborate the proposed 
scenarios for the origin of the UV upturn phenomenon. 

EHB stars in globular clusters have long been used to gain insight into the 
origin of the UV upturn phenomenon. However, the current evidence appears 
to increasingly point to different formation channels for field and cluster 
EHB stars~-- in particular, the binary fraction of field sdB stars appears 
to be very high, whereas (close) binary systems seem to be lacking in 
globular clusters. If confirmed by more extensive observations (only a 
couple of globular clusters have been adequately monitored so far), 
it may not be entirely appropriate 
anymore to rely on observations of resolved globulars to gain 
insight into the physical origin of the UV upturn phenomenon affecting 
elliptical galaxies and the bulges of spirals. Indeed, the evolutionary 
channels producing EHB stars, and hence the observed far-UV flux, may be 
quite different in the field and in clusters, binary stars plausibly 
playing a more important role in the case of the former.

\subsubsection{Acknowledgments}
The author is very grateful to the conference organizers for their 
assistance in presenting this paper, and to C. Moni Bidin for some 
helpful comments. This work is supported by Proyecto Fondecyt 
Regular No. 1071002. 

%%%%%%%%%%%%%%%%%%%%%%%%%%%%%%%%%%%%%%%%%%%%%%%%%%%%%%%%%%%%%%%%%%%

%%%%%%%%%%%%%%%%%%%%%%%%%%%%%%%%%%%%%%%%%%%%%%%%%%%%%%%%%%%%%%%%%%%%%%  }

\printindex
\end{document}